\documentclass[pre,aps,twocolumn,superscriptaddress,floatfix]{revtex4-1}

\usepackage{graphicx}
\usepackage{amsmath,amssymb}
\usepackage{color}

\begin{document}

\title{Diffusion properties of self-propelled particles in cellular flows}

\author{Lorenzo Caprini}
\affiliation{Gran Sasso Science Institute (GSSI), Via F.Crispi 7, I-67100 L'Aquila, Italy}

\author{Fabio Cecconi}
\affiliation{CNR-Istituto Sistemi Complessi, Via dei Taurini 19, I-00185, Rome, Italy}

\author{Andrea Puglisi}
\affiliation{CNR-Istituto Sistemi Complessi, P.le A. Moro, I-00185, Rome, Italy}

\author{Alessandro Sarracino}
\affiliation{Dipartimento di Ingegneria, Universit\`a della Campania ``L. Vanvitelli'', via Roma 29, 81031 Aversa (Caserta), Italy}

\begin{abstract}
  We study the dynamics of a self-propelled particle advected by a
  steady laminar flow.  The persistent motion of the self-propelled
  particle is described by an active Ornstein-Uhlenbeck process.  We
  focus on the diffusivity properties of the particle as a function of
  persistence time and free-diffusion coefficient, revealing
  non-monotonic behaviors, with the occurrence of a minimum and steep
  growth in the regime of large persistence time. In the latter limit,
  we obtain an analytical prediction for the scaling of the diffusion
  coefficient with the parameters of the active force. Our study sheds
  light on the effect of an inhomogeneous environment on the diffusion
  of active particles, such as living microorganisms and motile
  phytoplankton in fluids.
\end{abstract}

\maketitle


\section{Introduction}
In recent years, great interest has been raised by the complex
dynamics of biological microorganisms or artificial microswimmers,
which convert energy from the environment into systematic 
motion~\cite{bechinger2016active, marchetti2013hydrodynamics}.
Nowadays, these non-equilibrium systems are classified as
self-propelled particles (SPP), a subclass of dry active matter.  
The behavior of SPP without environmental constraints has been largely
studied and has shown a very fascinating phenomenology, such as: 
swarming, clustering~\cite{buttinoni2013dynamical, palacci2013living,
  ginot2018aggregation}, phase-separation~\cite{fily2012athermal,
  bialke2015active, cates2015motility, digregorio2018full,
  solon2018generalized, chiarantoni2020work}, spontaneous velocity
alignments~\cite{caprini2020spontaneous, lam2015self,
  ginelli2010large} and vortex formation~\cite{sumino2012large}.  
These phenomena have not a passive Brownian counterpart, because of the 
intrinsic non-equilibrium nature of the self-propelled motion, 
characterized by a persistence time
$\tau$, and, in particular, due to the peculiar interplay between active forces
and interactions among particles. Even in non-interacting cases, the
diffusive properties of SPP have been a matter of intense study both
experimentally and numerically since the self-propulsion enhances
the diffusivity with respect to any passive
tracers~\cite{ten2011brownian, sevilla2014theory, basu2018active,
scholz2018inertial, sevilla2019generalized, caprini2019activechiral}.

The dynamics of SPP in complex and non-homogeneous environments 
constitute a central issue for its great biological interest.
Indeed, in Nature, microswimmers or bacteria, when encounter soft or
solid obstacles~\cite{MinoClement2018EColi} or even hard walls~\cite{li2009accumulation}, accumulate in front of them producing interesting patterns \cite{wensink2008aggregation, Elgeti2013wall, wittmann2016active, das2020aggregation, caprini2019transport}. 
Morover, the swimming in porous soil~\cite{ford2007role}, blood
flow~\cite{engstler2007hydrodynamic} or biological
tissues~\cite{ribet2015bacterial} constitute other contexts of 
investigation.  
Thus, going beyond the description
of the active dynamics into homogeneous environments represents a
great challenge towards the comprehension of the life of microorganisms in 
their own habitat.  Moreover, the recent technological advances have led to the
possibility of manufacturing complex patterns of irregular or regular 
structures of micro-obstacles that
mimic the cellular environment or, more generally, the medium where
SPP move~\cite{chepizhko2013diffusion, chepizhko2013optimal,
  majmudar2012experiments, brown2016swimming, volpe2011microswimmers}.
The dynamics in complex environments has been studied for instance in
mazes~\cite{khatami2016active}, pinning
substrates~\cite{sandor2017dewetting}, arrays of
funnels~\cite{kaiser2012capture, tailleur2009sedimentation,
  wan2008rectification, volpe2014simulation, galajda2007wall}, comb
lattices~\cite{benichou2015diffusion},
fixed arrays of pillars~\cite{jakuszeit2019diffusion, pattanayak2019enhanced, reichhardt2018clogging, morin2017distortion} or even in
systems with random moving obstacles~\cite{zeitz2017active,
  aragones2018diffusion}, leading in some cases to anomalous
diffusion~\cite{barkai2012strange,shaebani2014anomalous, woillez2019active}.

Another important issue concerns the dynamics and diffusion properties
of SPP in environments which are non-homogeneous because of the
presence of a non-uniform velocity field.  In the case of passive
Brownian particles, the problem has been largely studied, both for
laminar and turbulent flows~\cite{shraiman1987diffusive}.  The
simplest example of the interplay between advection and molecular
diffusion is the Taylor dispersion~\cite{taylor1953dispersion} which
is observed in a channel with a Poiseuille flow.  In the case of SPP,
similar studies have great relevance in describing the behavior of
microscopic living organisms such as certain kinds of motile plankton
and microalgae.  For instance, the distribution of SPP in convective
fluxes is a problem that comes from the observations that plankton is
subject to Langmuir
circulation~\cite{thorpe2004langmuir,alonso2019transport}.  For
gyrotactic swimmers, such as certain motile phytoplanktons and
microalgae, the motion in the presence of flow fields, both in laminar
and turbulent regimes, has been studied
in~\cite{durham2013turbulence,santamaria2014gyrotactic}, showing that
strong heterogeneity in the distribution of particles can
occur. Interaction between laminar flow and motility has been studied
also in bacteria, with the observation of interesting trapping
phenomena~\cite{rusconi2014bacterial} and complex particle
trajectories~\cite{junot2019swimming}.

From the theoretical point of view, in the passive Brownian case, the
effective diffusion coefficient, $D_{\mathrm{eff}}$, of a tracer advected by a
laminar flow has been computed analytically in the limit of small
diffusivity by Shraiman~\cite{shraiman1987diffusive}. In this case,
$D_{\mathrm{eff}}$ is larger than the free-diffusion coefficient $D_0$ (the
diffusivity in the absence of convective flow).  A first-order correction
accounting for the noise persistence has been computed by Castiglione
and Crisanti~\cite{castiglione1999dispersion}, who estimated
asymptotically $D_{\mathrm{eff}}$ for vanishing persistence time,
finding a further enhancement of diffusivity. 

In this manuscript, we generalize the study of the diffusion of SPP in
the presence of laminar flows, for large values of the persistence time 
$\tau$.  The effect
of an underlying cellular flow has been considered for instance
in~\cite{torney2007transport}, revealing non-trivial effects such as
negative differential and absolute mobility when the particles are
subject to an external force~\cite{sarracino2016nonlinear,
  cecconi2017anomalous,cecconi2018anomalous}.  At variance
with~\cite{torney2007transport}, where concentration in some flow
regions was investigated, we focus here on the transport properties,
such as the diffusion coefficient and the mean square displacement.
Our analysis unveils a rich phenomenology, characterized by a
nonmonotonic behavior of $D_{\mathrm{eff}}$ as a function of the persistence
time of the active force, which implies that, in some cases, the
coupling of the active force with the underlying velocity field can
trap the SPP, resulting in a decrease of diffusivity.  The dynamics of
active particles in convective rolls has recently been considered
in~\cite{li2020diffusion} where the authors focus on the role of
chirality.

The paper is structured as follows: after the introduction of the
model in Sec.~\ref{sec:Model}, we present our numerical results in
Sec.\ref{sec:numericalresults}, focusing on the mean square
displacement and effective diffusion coefficient as a function of the
model parameters.  In Sec.\ref{sec:analytical}, we present an
analytical computation which explains the behavior of $D_{\mathrm{eff}}$ in the 
large persistence regime. 
Then, we summarize the result in the conclusive section.


\section{The Model \label{sec:Model}}
We consider a dilute system of active particles in two dimensions
diffusing in a cellular flow.  The position, $\mathbf{r}=(x,y)$, of
the tagged SPP is described by the following stochastic differential
equation:
\begin{equation}
\label{eq:xdynamics}
\dot{\mathbf{r}} = \mathbf{A}(\mathbf r) + \mathbf{w} \,,
\end{equation}
where we have neglected the thermal Brownian motion due to the solvent
as also any inertial effects, as usual in these systems~\cite{bechinger2016active}.  The
self-propulsion mechanism is represented by the term $\mathbf{w}$,
evolving according to the Active Ornstein-Uhlenbeck particle (AOUP)
dynamics.  AOUP is an established model to describe the behavior of
SPP~\cite{das2018confined, fodor2016far, caprini2019entropy,
  wittmann2019pressure, berthier2019glassy, woillez2019nonlocal,
  marconi2016velocity} or passive particles immersed into an active
bath~\cite{maggi2014generalized, maggi2017memory}.  The colored noise,
$\mathbf{w}$, models the persistent motion of a SPP conferring 
time-persistence to a single trajectory and evolves
according to the equation
\begin{equation}
\label{eq:wdynamics}
\tau\dot{\mathbf{w}}  = -\mathbf{w} + \sqrt{2D_0}\;\boldsymbol{\xi} \,,
\end{equation}
where $\tau$ is the auto-correlation time of $\mathbf{w}$ and sets the
persistence of the dynamics, while the constant $D_0$ represents the
effective diffusion due to the self-propulsion in a homogeneous environment.
The ratio $\sqrt{D_0/\tau}$ determines the average velocity induced by
the self-propulsion, being $D_0/\tau$ the variance of $\mathbf{w}$.

The cellular flow, $\mathbf{A}$, is chosen as a periodic,
divergenceless field obtained from the stream function
\begin{equation}\label{stream}
\psi(\mathbf r) = \dfrac{U_0}{k} \sin(k x) \sin(k y) \,,
\end{equation}
where $U_0$ sets the maximal intensity of the field, while $k=2\pi/L$
determines the cell periodicity, with $L$ the cell size.  The flow is
obtained from the stream function as
\begin{equation} 
(A_x,A_y) = \{\partial_y \psi(\mathbf r), -\partial_x \psi(\mathbf r)\} \,.
\end{equation}
Basically, the flow is a square lattice of convective cells (vortices)
with alternated directions of rotation. The boundary lines separating
neighboring cells are called ``separatrices'': along a separatrix, the
flow has a maximal velocity in the parallel direction and zero in the
perpendicular one.  The structure of the cellular flow introduces a
time-scale in the dynamics of the system, the turnover time $T_U =
L/U_0$, i.e. the time needed by a particle, in the absence of any
other forces, to explore the whole periodicity of the system.  The
self-propulsion is characterized by the typical time $\tau$, whose
interplay with $T_U$ determines a complex phenomenology, both at the
level of a single particle trajectory and at the diffusive level, as
it will be illustrated in the next sections.

We remark that a self-propelled particle immersed in flow cannot be studied
employing any suitable approximations, such as the Unified Colored
Noise Approximation~\cite{maggi2015multidimensional,
  caprini2019activity}, except for the small persistence regime,
defined for values of $\tau$ such that $\tau<\tau^*=1/kU_0= T_U/2\pi$ , as shown
in Appendix~\ref{Appendix:UCNAfailure}. Indeed, the form of the
velocity dynamics in the large persistence regime
prevents the possibility of
adapting the adiabatic elimination, except in the limit of small $\tau$~\cite{caprini2019active}.



\section{Numerical Results}\label{sec:numericalresults}

\begin{figure}[!t]
\centering
\includegraphics[clip=true,width=0.9\columnwidth,keepaspectratio]
{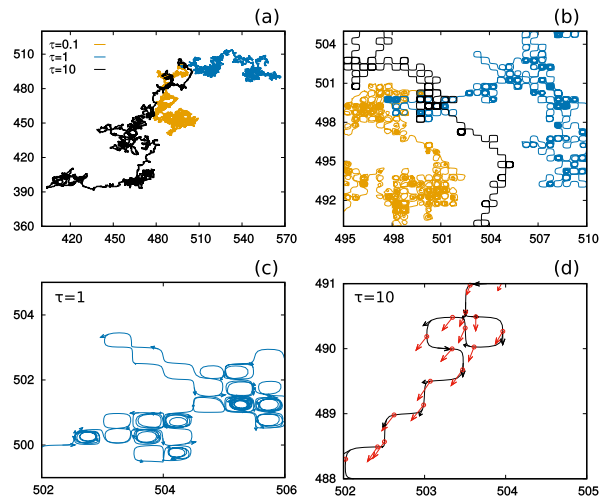}
\caption{(a)-(b) panels: Snapshot of the trajectories for different
  values of $\tau=0.1, 1, 10$. (c)-(d) panels: enlargement for
  $\tau=1$ (c) and for $\tau=10$ (d). In panel (c) the arrows
  represent the velocity of the particle. In panel (c), the black
  arrows denote the velocity, while the red arrows denote the
  self-propulsion. The self-propulsion is rescaled for presentation
  reasons but here is $\sim 10^{-2}$ smaller than $U_0$.  Simulations are
  realized with $D_0=10^{-2}$.  The other parameters are $U_0=1$ and
  $L=1$.  }
\label{fig:Snapshot}
\end{figure}

\begin{figure*}[t]
\centering
\includegraphics[width=1\textwidth,clip=true]
{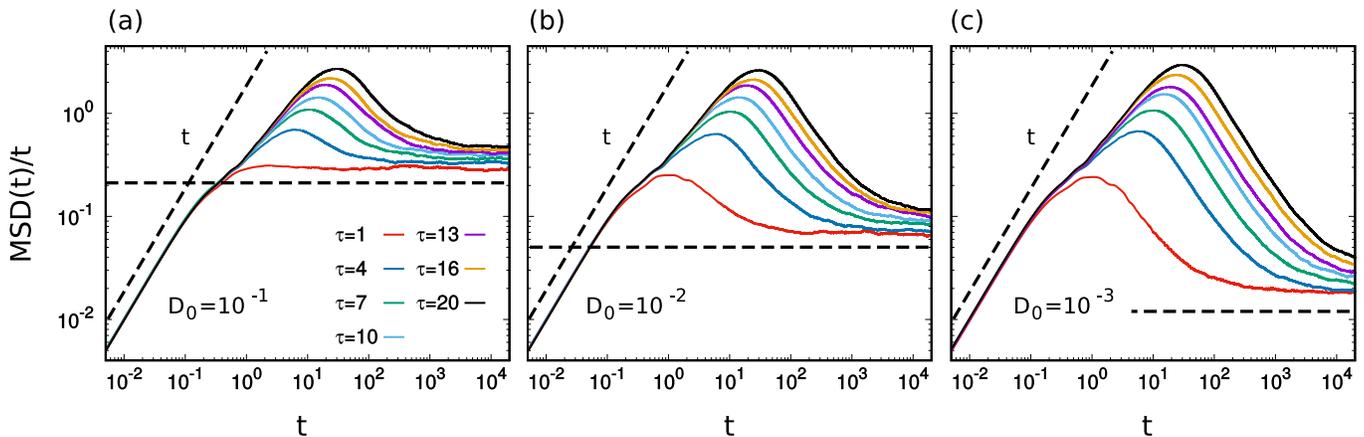}
\caption{$\mathrm{MSD}(t)/t$ as a function of $t$ for different values of
  $\tau$, namely $\tau=1, 4, 7, 10, 13, 16, 20$ from the bottom to the
  top as shown in the legend. Panels (a), (b) and (c) are obtained
  with three different values of $D_0=10^{-1}, 10^{-2}, 10^{-3}$ from
  the left to the right.  Dashed black lines mark linear and constant
  behaviors (corresponding on the ballistic and diffusive regimes of
  the $\mathrm{MSD}(t)$, respectively).  The other parameters are $U_0=1$ and
  $L=1$. 
}
\label{fig3}
\end{figure*}

We carry out a numerical study of the active
dynamics~\eqref{eq:xdynamics} and~\eqref{eq:wdynamics} that are integrated 
via a second order stochastic Runge-Kutta algorithm \cite{RK2ord} with a time step $h = 10^{-3}$ and for a time at least $2\times10^2\,\tau$.  
Simulations are performed keeping fixed the cellular structure, 
$L=1$, $U_0=1$, in such a way that $T_U=1$, and evaluating the influence 
of the parameters $\tau$ and $D_0$ on the dynamics. 


\subsection{Single particle trajectories}

We start by studying qualitatively the typical trajectories of the SPP
in different relevant regimes. These observations will help us to
understand the average properties showed by the mean square
displacement and by the diffusion coefficient. In particular, in
Fig.~\ref{fig:Snapshot} (a), we compare different single-particle
trajectories obtained for three different values of $\tau=10^{-1},1,10$
at fixed $D_0=10^{-2}$.  For $\tau \ll T_U$ (yellow trajectory), the
persistence feature of the self-propulsion is not relevant since
$\mathbf{w}$ changes direction many times inside a single cell.  As a
consequence, the self-propulsion is indistinguishable from a thermal
one and $\mathbf{w}$ behaves as a thermal noise with diffusivity
$D_0$ (see Sec.~\ref{sec:numericalresults}~C).

For $\tau \gg T_U$ (black trajectory), the self-propulsion 
changes only after that the particle has crossed many
cells. In these regimes, the average speed of self-propulsion
is very small, decreasing as $1/\sqrt{\tau}$, since $D_0$ is fixed.  As shown
in panels (b) and (d) of Fig.~\ref{fig:Snapshot}, the particle
proceeds along the separatrix between different vortices, and explores
the regions where the cellular flow assumes its maximal value.  The
particle does not explore the region inside the cell and quite rarely
becomes trapped in a vortex. This event is rarer as $\tau$ is
increased.  As a consequence, the self-propelled particle moves in a
zig-zag-like way with a trajectory displaying an almost deterministic
behavior that follows the flow field as shown in
Fig.~\ref{fig:Snapshot} (d).  Due to the small value of the self-propulsion compared to $U_0$,
$\mathbf{w}$ plays a role only in a small region near the nodes of the
separatrix (where the flow field is zero). In those regions, the
direction of $\mathbf{w}$ determines which one of the two separatrices
the particle will follow.  Moreover, since the average change of the
self-propulsion direction is ruled by $\tau$, the same separatrix will
be preferred for times smaller than $\tau$.  This results in a unidirectional motion
(along the separatrices) for small times ($< \tau$), in analogy with a free active
particle with velocity $U_0$, while a diffusive-like behavior will be
obtained for times larger than $T_U $.

For intermediate values of $\tau$, i.e. when $\tau\sim T_U$, the
trajectory is more complicated, as illustrated in
Fig.\ref{fig:Snapshot} (c).  The self-propulsion can deviate the
trajectory from the cellular flow and push the SPP inside a vortex.
The exit from the vortex can be determined by a fluctuation of the
self-propulsion.  A particle needs more time with respect to the
thermal case to escape and proceed along the separatrix.

\subsection{Mean Square Displacement}
\label{sec:diffusiveproperties}
The rescaled mean square displacement (MSD) of the SPP, $\langle
[x(t)-x(0)]^2\rangle/t$, averaged over thousands of realizations, is
reported for several values of $\tau$ and three values of $D_0$ (Fig.~\ref{fig3}~(a)-(c)).
The $\mathrm{MSD}(t)$ 
reflects the behaviors of the single-particle trajectories: we identify
short-time ballistic, intermediate-time anomalous diffusive and long-time diffusive regimes.  
In the small-$\tau$ limit, such that $\tau<\tau^*$, ballistic regimes occur for $t<\tau$ (not shown in the figure), in analogy with active particles in a homogeneous environment~\cite{caprini2019activechiral}.
When $\tau > \tau^*$,  deviations from ballistic regimes occur for $t>\tau^*$, as reported in Fig.~\ref{fig3}~(a)-(c).
As shown in each panel, this regime weakly depends
on $\tau$ and $D_0$ since for small $t$ the MSD collapses onto the same
curve, at variance with active particles in homogeneous environments. 
A second regime occurs in the range of times $\tau^* < t <\tau$ where
diffusion is slower but still superdiffusive, until a maximum of
$\langle [x(t)-x(0)]^2\rangle/t$ is observed for $t \sim \tau$. After
this maximum, the diffusivity slows down (the crossover appears as a
transient sub-diffusion) and gets to normal diffusion asymptotically.
The comparison between the different panels of Fig.~\ref{fig3}
suggests that the anomalous diffusive regimes are less pronounced as
$D_0$ increases even if, in all the cases, the anomalous diffusion
region enlarges as $\tau$ grows.  At large times, after $t>
\tau_D$, the diffusive behavior is reached.  Remarkably, the transient
regime has a quite long duration, and the typical time $\tau_D$
increases with $\tau$ and decreases with $D_0$.  We remark that the
slowing down of the dynamics at intermediate times is related to the
trapping effect due to the vortices of the cellular flow, which
confines the particle motion in a limited region for a certain time.



\subsection{Diffusion coefficient}

\begin{figure}[t]
\centering
\includegraphics[width=0.9\columnwidth,clip=true]{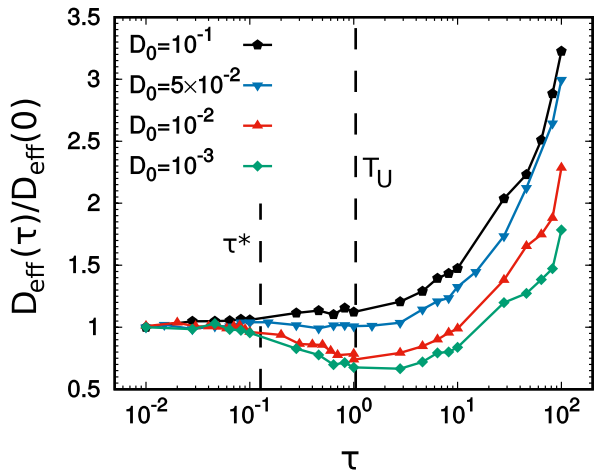}
\caption{Effective diffusion coefficient as a function of $\tau$, for several
  values of $D_0$ as shown in the legend. The two dashed black lines are eye-guides which mark the value of $\tau^*$ and $T_U$.
The other parameters are $U_0=1$ and $L=1$.
}
\label{fig1}
\end{figure}

To unveil the effect of the self-propulsion force on the long-time
diffusive dynamics, we study the diffusion coefficient
\begin{equation}
D_{\mathrm{eff}}=\lim_{t\to\infty}\frac{1}{2t}\langle [x(t)-x(0)]^2\rangle \,,
  \end{equation}
as a function of the activity parameters, $\tau$ and $D_0$.  The case
$\tau=0$ corresponds to the passive Brownian limit, where the leading
contribution to the diffusion comes from the particles which move on
the separatrices, and for which an analytical prediction has been
computed by Shraiman in~\cite{shraiman1987diffusive}:
\begin{equation}
D_{\mathrm{eff}}(\tau=0) = 
\frac{S(k=1)}{\sqrt{\pi}} \sqrt{\dfrac{U_0 L D_0}{2\pi}}\,.
  \label{sh}
\end{equation}
Here, the function $S(k)$ depends on the cell geometry and is reported
in Ref.~\cite{shraiman1987diffusive}.  For the setup employed in the
numerical study of this manuscript, we have $S(k=1)/\sqrt{\pi}\simeq
1.07$.  We remark that the cellular flow enhances the diffusivity at
small $D_0$, with respect to the case of homogeneous environments,
producing a scaling $\sim\sqrt{D_0}$ instead of $\sim
D_0$.

\begin{figure}[t]
\centering
\includegraphics[width=0.9\columnwidth,clip=true]{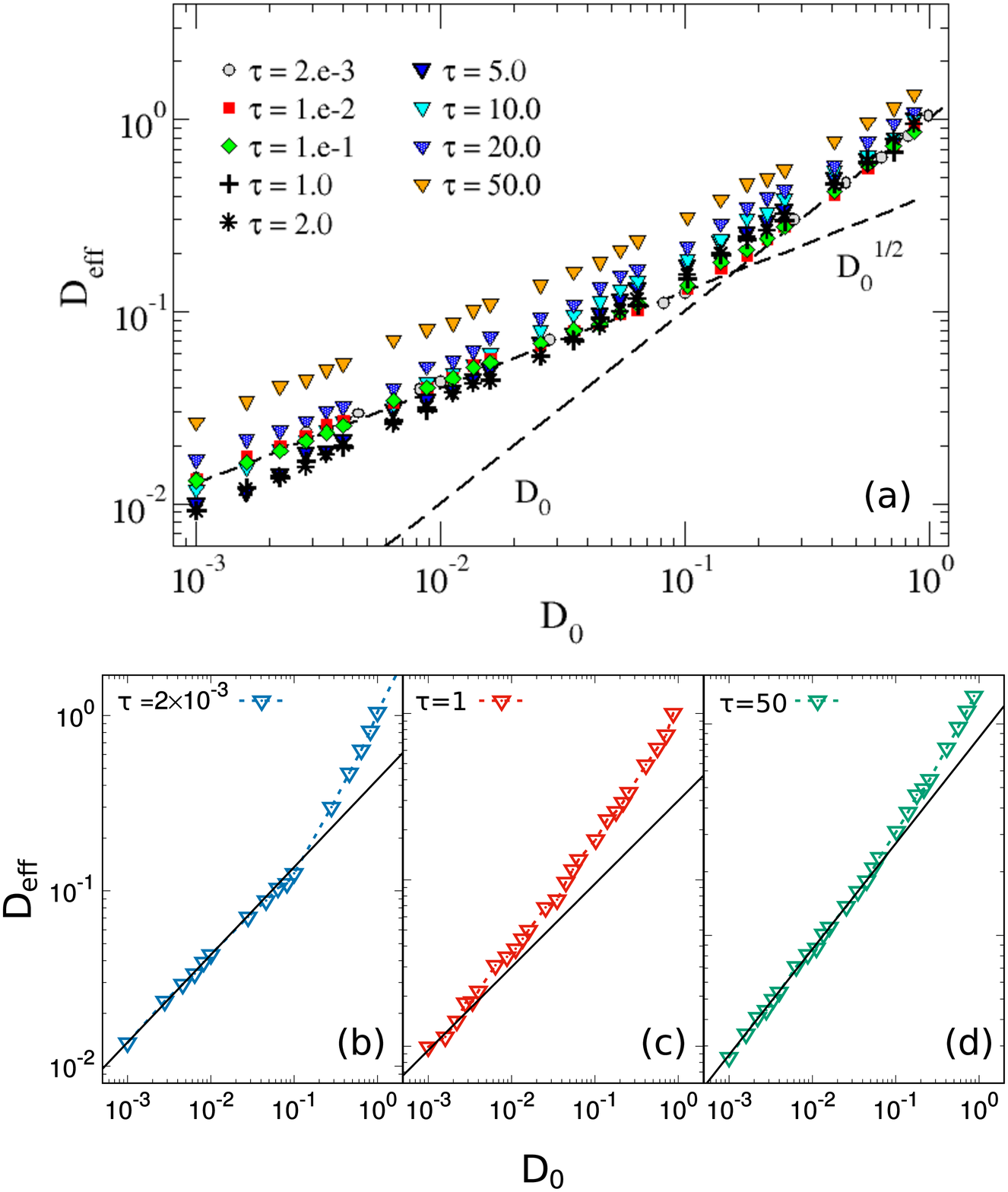}
\caption{Effective diffusion coefficient, $D_{\mathrm{eff}}$, vs. $D_0$ for
  different values of $\tau$, as shown in the legend (panel (a)).
The dashed black lines are eye guides showing the behavior $\sim D_0$ and $\sim \sqrt{D_0}$.
  Panels (b), (c) and (d) compare a curve of $D_{\mathrm{eff}}$ (for a given
  $\tau$) with the behavior $\propto \sqrt{D_0}$.
The other parameters are $U_0=1$ and $L=1$.
}
\label{fig2}
\end{figure}

In Fig.~\ref{fig1}, we plot
$D_{\mathrm{eff}}(\tau)/D_{\mathrm{eff}}(0)$ as a function of $\tau$
for four values of $D_0$ to show the role of the
self-propulsion persistence.  As expected, at small values of $\tau$,
the prediction~(\ref{sh}) is in agreement with numerical simulations
since the self-propulsion acts as an effective thermal noise, in this
regime.  Depending on the value of $D_0$, we observe a different
phenomenology.  In particular, for the larger values of $D_0$, for
instance $D_0=10^{-1}$ (or larger), $D_{\mathrm{eff}}$ grows with $\tau$ and,
thus, the effect of increasing the persistence time is to enhance the
diffusivity, even if the effective velocity decreases as
$\sqrt{D_0/\tau}$.  More surprisingly, for the smaller values of
$D_0$, we obtain a non-monotonic behavior: In a regime of $\tau$
comparable with $T_U$, starting from $\tau^* = 1/kU_0$, we observe
that $D_{\mathrm{eff}}$ decreases down to a minimum value which is reached at times close to $T_U$.  We remark that $\tau^*$ is the value
of $\tau$ for which the overdamped approach for small $\tau$ does not
hold, as shown in detail in Appendix~\ref{Appendix:UCNAfailure}.
After the minimum is reached, $D_{\mathrm{eff}}$ grows indefinitely.  In particular, for
$\tau$ large enough we observe $D_{\mathrm{eff}}> D_{\mathrm{eff}}(\tau=0)$ as in cases with larger $D_0$.

These observations are in agreement with the phenomenology characterizing the
single-particle trajectories (Fig.~\ref{fig:Snapshot}). Indeed, the
possibility of being trapped into a vortex for long times - seen for
values of $\tau \sim T_U$ - is coherent with the observed reduction of
$D_{\mathrm{eff}}$ in a range of $\tau$.  Also, the observation of trajectories
running fast along the separatrices for large values of $\tau$ is
compatible with the final growth of $D_{\mathrm{eff}}$ (asymptotically for
large $\tau$). Using this information, in the next section, we will
derive an analytical prediction for $D_{\mathrm{eff}}$, in the regime $\tau \gg
T_U$.


In panel~(a) of Fig.~\ref{fig2}, we show $D_{\mathrm{eff}}$ as a
function of $D_0$ for several values of $\tau$ to test Shraiman's
scaling with $D_0$.  As shown in panel (b) of Fig.~\ref{fig2}, for
$\tau\ll T_U$, $D_{\mathrm{eff}}\sim \sqrt{D_0}$, in agreement with
Eq.~\eqref{sh}.  Shraiman's scaling breaks down for large values of
$D_0$, where $D_{\mathrm{eff}} \propto D_0$, occurring when  $\mathbf{w}$ becomes comparable
with $U_0$ and the cellular-flow plays a marginal role in the
transport process.  In Fig.~\ref{fig2} (c), we observe that Shraiman's
scaling does not hold for the intermediate values of $\tau$ (the
region of $\tau$ corresponding to the minimum in Fig.~\ref{fig1}),
while it is recovered in the regime of large $\tau$, namely for
$\tau\gg T_U$, even if the values of $D_{\mathrm{eff}}$ are quite larger with
respect to Eq.~\eqref{sh}, see panel~(d).

\section{Analytical prediction of $D_{\mathrm{eff}}$ for large persistence
\label{sec:analytical}}

\begin{figure}[t]
\centering
\includegraphics[width=0.9\columnwidth,clip=true]{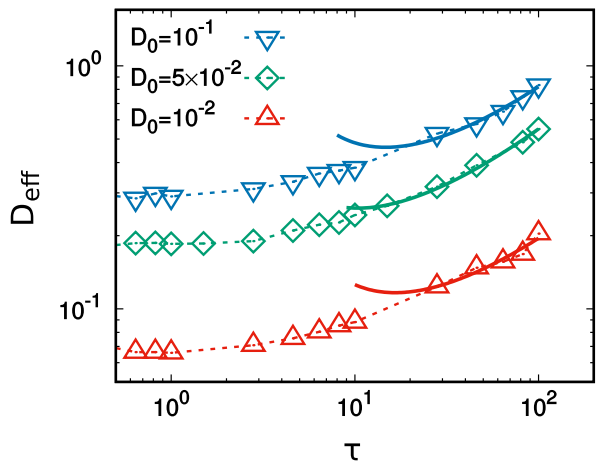}
\caption{Effective diffusion coefficient vs. $\tau$ for two different
  values of $D_0$ (colored data). 
The solid lines are obtained from numerical fits of the prediction~\eqref{eq:prediction_Deff_largetau}, 
namely $g(\tau) = a*\tau / \log^2( 2 U_0^2 \tau/D_0 /b)$, where $a$ and $b$ are two parameters.
In particular, $b \approx 6$ does not depend on $D_0$ and $T_U$ and is just a numerical factor.  
The parameters of the numerical study are $U_0=1$ and $L=1$.  
}
\label{fig:prediction}
\end{figure}

In the large persistence regime, $\tau \gg T_U$, the study of the
single-particle trajectory has revealed that the SPP runs almost
deterministically along the separatrices choosing the ``same''
direction for a time of order $\tau$. The change of direction occurs after a
time $\sim\tau$, as it happens for an active particle
in a homogeneous environment. 
In this simple case, the MSD, at large times, is given by
\begin{equation}
\mathrm{MSD}(t\gg \tau)  \approx {\mathcal L}^2 n=\left(\sqrt{D_0\tau}\right)^2 \frac{t}{\tau} \,,
\end{equation}
where ${\mathcal L}=\sqrt{D_0\tau}$ is the persistence length and
$n=t/\tau$ counts the number of persistence lengths covered by the
free-particle in the time $t$.

We consider the MSD of a self-propelled particle in the absence of cellular
flow and replace the persistence length in a homogeneous environment
${\mathcal L}$ with the persistence length ${\mathcal L}_{\mathrm{eff}}$ of a
particle moving in the laminar flow along the separatrices, with velocity $U_0 \sin{\left(kx\right)}$ for
$x \in [0, L/2]$ (obtained from Eq.~(\ref{eq:xdynamics})).  Thus, at
large times, we get the estimate
\begin{equation}
\label{eq:predictionDeff}
\mathrm{MSD}(t\gg \tau) \approx {\mathcal L}_{\mathrm{eff}}^2 \frac{t}{\tau}= \left(\frac{L}{2\sqrt{2}}\frac{\tau}{t_U}\right)^2 \frac{t}{\tau} \,,
\end{equation}
where $t_U$ is the typical time to run for $L/2$ along a separatrix
and is calculated in
Appendix~\ref{Appendix:A}.
We remark that the validity of Eq.~\eqref{eq:predictionDeff} 
is restricted to $\tau \gg T_U$, a regime where 
$\mathbf{w}$ plays a role only on the nodes of the separatrices because
$|\mathbf{w}|<U_0$. 


This argument suggests that the diffusion coefficient increases
linearly with $\tau$ and does not depend on $D_0$, except for a
dependence contained in $t_U$.  In particular, we get (see 
Appendix~\ref{Appendix:A})
\begin{equation}
\label{eq:TUmain}
t_U\approx \frac{L}{U_0 2\pi}\log{\left[ \frac{2}{b}\frac{\tau}{D_0}{U_0^2}    \right]} \,,
\end{equation}
in the limit $\sqrt{D_0}/\sqrt{\tau} \ll U_0$.
The parameter $b$ is just a numerical factor which does not depend on $\tau$, $D_0$ and $T_U$. 
In this regime, the prediction for the diffusion coefficient, for $\tau \gg
t_U$ (calculated for $t \gg \tau$), reads:
\begin{equation}
\label{eq:prediction_Deff_largetau}
D_{\mathrm{eff}} \propto \dfrac{U_0^2\,\tau}{\log^2{\left[ \frac{2}{b}{U_0^2}\frac{\tau}{D_0}    \right]} } \,.
\end{equation}
The comparison between prediction and numerical data is reported in
Fig.~\ref{fig:prediction} as a function of $\tau$ for three different
values of $D_0$.
The results are in good agreement for $\tau \gg T_U$, while  
marked deviations emerge for $\tau \sim T_U$, where the main hypothesis behind
Eq.~\eqref{eq:prediction_Deff_largetau} does not apply.

\section{Conclusion}

In this manuscript, we have studied the diffusive properties of a
self-propelled particle moving in a steady laminar flow, evaluating
the effect of the self-propulsion.  
The diffusion coefficient displays a non-monotonic behavior as a function of the persistence
time $\tau$: in particular, a minimum occurs for a large range of
$D_0$ when $\tau$ is comparable with the turnover time, followed by a
sharp increase, faster than $\tau$, such that the value of the
diffusion coefficient exceeds Shraiman's prediction valid in the
passive Brownian case.  Such a mechanism is discussed and connected to
the single-particle trajectory, specifically to the occurrence
of a trapping mechanism into the vortices.  Additionally, Shraiman's
scaling with the diffusion coefficient is tested in the active
case, revealing an intriguing scenario.

Our study shows that the presence of the self-propulsion affects 
the diffusion in a complex environment and could
represent a mechanism naturally developed by self-propelled agents to
improve the efficiency of the transport. Testing the presence of
similar nonmonotonic behaviors in other inhomogeneous environments,
going beyond the specific functional form of a laminar flow field,
could be a promising research line to understand the behavior of
self-propelled microorganisms in their complex habitats.

\section*{Acknowledgements}
{L. Caprini, F. Cecconi, A. Puglisi and A. Sarracino
  acknowledge support from the MIUR PRIN 2017 project
  201798CZLJ. A. Sarracino acknowledges support from Program
  (VAnviteLli pEr la RicErca: VALERE) 2019 financed by the Univeristy
  of Campania ``L. Vanvitelli''.
  F. Cecconi and A. Puglisi acknowledge the financial support of Regione Lazio through the 
  Grant ``Progetti Gruppi di Ricerca'' N. 85-2017-15257.
}

\appendix

\section{Failure of the UCNA approximation 
\label{Appendix:UCNAfailure}}
By taking the time derivative of the equation of
motion~(\ref{eq:xdynamics}), we get:
\begin{flalign}
\ddot{\mathbf{r}} & = \nabla \mathbf{A}(\mathbf{r})\;
\dot{\mathbf{r}} + \dot{\mathbf{w}}\\
\tau\dot{\mathbf{w}} & =  -\mathbf{w} + \sqrt{2D_0}\;\boldsymbol{\xi}\,,
\end{flalign}
where the matrix $ \nabla \mathbf{A}$ reads:
\begin{equation*}
\begin{aligned}
\nabla \mathbf{A} &= kU_0
\begin{bmatrix} 
\cos{\left(kx\right)} \cos{\left(ky\right)} \quad& -\sin{\left(kx\right)} \sin{\left(ky\right)}\\\\
\sin{\left(kx\right)} \sin{\left(ky\right)} \quad& -\cos{\left(kx\right)}\cos{\left(ky\right)} 
\end{bmatrix} \\
&=k^2
\left[
\begin{array}{cc}
\phi~ &  -\psi \\
\psi~ &  -\phi
\end{array}
\right]\,,
\end{aligned}
\end{equation*}
being $\psi$ the stream function defined by Eq.~\eqref{stream} and 
$$
\phi(\mathbf{r})=\frac{U_0}{k} \cos{(kx)}\cos{(ky)}\,.
$$
Adopting the usual change of variable 
$\mathbf{v} = \dot{\mathbf{r}}$, $\mathbf{v} = \mathbf{A}(\mathbf r) + \mathbf{w}$, we obtain
\begin{equation}
\label{eq:transformedEqofmotion}
\tau\dot{\mathbf{v}} = \boldsymbol{\Gamma} \mathbf{v}  + \mathbf{A}(\mathbf{r}) +  \sqrt{2D_0}\boldsymbol{\xi} \,,
\end{equation} 
where the matrix $\boldsymbol{\Gamma}$ assumes the simple form:
\begin{equation}
\label{eq:app_Gamma}
\boldsymbol{\Gamma}(\mathbf{r}) = 
\mathcal{I} - \tau\nabla \mathbf{A}(\mathbf{r})\,,
\end{equation}
and $\mathcal{I}$ is the identity matrix.
For those values of $\tau$ such that the matrix 
$\boldsymbol{\Gamma}$ is no longer positive-defined,  
the overdamped limit needed for UCNA becomes meaningless. 
We recall that a diagonalizable matrix is positive-defined when all its eigenvalues are 
positive. 
The eigenvalues of the matrix $\boldsymbol{\Gamma}$ (Eq.~\eqref{eq:app_Gamma}) read:
\begin{flalign*}
\lambda_1 &= 1 - \tau k^2\sqrt{\phi^2 -\psi^2}  \\
\lambda_2 &= 1 + \tau k^2\sqrt{\phi^2 -\psi^2} \,.
\end{flalign*}
Using the definition of $\psi(x,y)$ and $\phi(x,y)$, after some algebraic 
manipulations, we obtain 
\begin{flalign*}
\lambda_1 &= 1 - \tau k U_0\sqrt{\cos[k(x-y)]\cos[k(x+y)]} \\
\lambda_2 &= 1 + \tau k U_0\sqrt{\cos[k(x-y)]\cos[k(x+y)]}
\end{flalign*}
When $\tau \geq \tau^* = 1/(U_0 k)$, the eigenvalue $\lambda_1$ has no 
possibility to be positive globally in space, then $\boldsymbol{\Gamma}$ cannot be positive definte and the overdamped regime turns to be undefined. 
In the opposite 
regime, $\tau <\tau^*$, which we call small-$\tau$ limit, the
overdamped regime could be assumed. 
Only, in the latter case, the
dynamics can be recast onto:
\begin{equation}
 \boldsymbol{\Gamma}\,{\mathbf v} = {\mathbf A}({\mathbf r}) 
+ \sqrt{2D_0}\;{\mathbf \xi}\,.
\end{equation} 
By inversion we obtain: 
\begin{equation}
\label{eq:app_UCNA}
\dot{\mathbf{r}}={\mathbf v} = 
\boldsymbol{\Gamma}^{-1}\bigg({\mathbf A}({\mathbf r)} + \sqrt{2D_0}\;
{\mathbf \xi}\bigg) \,,
\end{equation}
which corresponds to the UCNA dynamics~\cite{maggi2015multidimensional} adapted to the current case.

We remark that $\tau^*$ roughly corresponds to the value of $\tau$ at which
$D_{\mathrm{eff}}$ starts to consistently change with respect to the Shraiman's
prediction, as shown in Fig.~\ref{fig1}.

\section{Computation of $t_U$ in the regime of large $\tau$ 
\label{Appendix:A}}
The typical time $t_U$ contained in 
Eqs.~\eqref{eq:predictionDeff} and~\eqref{eq:TUmain} can be obtained by 
integrating Eq.~\eqref{eq:xdynamics}  without the self-propulsion in a given
direction along a separatrix, for instance:
\begin{equation}
t_U = \frac{1}{U_0} \int_0^{L/2} \frac{dx}{\sin{(2\pi x/L)}} \,.
\end{equation}
This procedure is justified because the active force is negligible along the separatrices, except for the nodes, in the large persistence regime.
The integral defining $t_U$ is not converging, unless we introduce two cut-offs
\begin{equation}
\begin{aligned}
t_U &= \frac{1}{U_0} \int_{x_m}^{L/2-x_m} \frac{dx}{\sin{(2\pi x/L)}} 
\,.
\end{aligned}
\end{equation}
The length scale $x_m$ is chosen such as:
\begin{equation}
\label{eq:app_xm}
U_0 \sin{\left(k x_m\right)} = \sqrt{2\,b\, \frac{D_0}{\tau}} \,,
\end{equation}
i.e. when the force along a separatrix is roughly equal to the typical value of the 
activity, estimated by its standard deviation, $\sqrt{D_0/\tau}$.
The factor $b$ is a parameter that does not depend on $\tau$, $D_0$ and $T_U$, being just a numerical factor.
Inverting Eq.~\eqref{eq:app_xm}, we get
\begin{equation*}
\begin{aligned}
x_m = \frac{L}{2\pi} \arcsin{\left(\sqrt{\frac{2 \,b\,D_0}{\tau}}\frac{1}{U_0}\right)} \approx  \frac{L}{U_0\pi} \sqrt{\frac{b\,D_0}{2\tau}}  \,,
\end{aligned}
\end{equation*}
where we have used the condition $\sqrt{D_0/\tau} \ll U_0=1$ holding in the large persistence regime.
Solving the integral, we obtain the final expression for $t_U$, 
\begin{equation*}
\label{eq:app_finaleqtu}
\begin{aligned}
  t_U\frac{U_0 \pi}{L}=& \log{\left( \frac{1}{\tan{\left[ \sqrt{\frac{b\,D_0}{2\tau}}\frac{1}{U_0}\right]} } \right)}
\approx \frac{1}{2}\log{\left( \frac{2 U^2_0}{b} \frac{\tau}{D_0}  \right)} \,,
\end{aligned}
\end{equation*}
which is positive since $U_0 \sqrt{\tau /D_0} \gg 1$.
	Thus, from Eq.~\eqref{eq:app_finaleqtu}, $t_U$ contains a dependence on $D_0/\tau$ which scales as $\log{(\tau/D_0)}$, as reported in Eq.~\eqref{eq:TUmain}.

\bibliographystyle{apsrev4-1}

\bibliography{activeMIPS.bib}

\end{document}